# A Parallel Optimal Task Allocation Mechanism for Large-Scale Mobile Edge Computing

Xiaoxiong Zhong, *Member, IEEE*, Xinghan Wang, Yuanyuan Yang, *Fellow, IEEE*,

Yang Qin, *senior Member, IEEE*, Xiaoke Ma, Tingting Yang, *Member, IEEE*

*Abstract-* We consider the problem of intelligent and efficient task allocation mechanism in large-scale mobile edge computing (MEC), which can reduce delay and energy consumption in a parallel and distributed optimization. In this paper, we study the joint optimization model to consider cooperative task management mechanism among mobile terminals (MT), macro cell base station (MBS), and multiple small cell base station (SBS) for large-scale MEC applications. We propose a parallel multi-block <u>A</u>lternating <u>D</u>irection <u>M</u>ethod of <u>M</u>ultipliers (ADMM) based method to model both requirements of low delay and low energy consumption in the MEC system which formulates the task allocation under those requirements as a nonlinear 0-1 integer programming problem. To solve the optimization problem, we develop an efficient combination of conjugate gradient, Newton and linear search techniques based algorithm with Logarithmic Smoothing (for global variables updating) and the <u>C</u>yclic <u>B</u>lock coordinate <u>G</u>radient <u>P</u>rojection (CBGP, for local variables updating) methods, which can guarantee convergence and reduce computational complexity with a good scalability. Numerical results demonstrate the effectiveness of the proposed mechanism and it can effectively reduce delay and energy consumption for a large-scale MEC system.

## I. INTRODUCTION

With the rapid development of the wireless technology and Internet, more and more mobile terminals (MT) have different wireless access requirements for bandwidth and computing, which promotes mobile applications, such as online playing game, virtual reality, intelligent data processing and other new services continue to emerge [1-3]. However, these new mobile applications have high energy consumption and high latency, which pose a huge challenge for computing and battery capacity of the MT. Mobile Edge Computing (MEC) is a new promotion technology that supports cloud computing capabilities and edge service environments at the edge of cellular networks. The legacy base station (BS) is updated to a MEC-enabled base station (MEC-BS) by being equipped with a computing function (such as an MEC server), hence the MEC-BSs can implement MT capability enhancement, which can reduce MT application execution time and MT energy consumption. The development of these new applications and services is limited by the computing capability and battery of these MTs. If data consumption and computationally intensive tasks need to be offloaded to the cloud for execution, it can address the problem that MTs' have a lack of computing capability. However, a relatively large delay will be induced when an MT connects to the cloud over a wireless network, which is not suitable for delay-sensitive tasks.

MEC allows MT to perform computational offloading to offload its computational tasks to the MEC-BS that overwrites it. When the execution of the task at the MEC-BS is completed, the MEC-BS returns the result of the task to the MT. Due to limited computing resources, MEC-BS is unable to provide unlimited computational offload services for all tasks in the MT within coverage. Therefore, how to effectively manage MEC-BS resources (e.g., energy and computational resource), maximizing system performance, is critical. On the other hand, if there are too many tasks to be uninstalled from the MT, the MEC-BS will still be overloaded. Hence, how to design an efficient task management scheme, which can reduce delay and energy consumption, is a challenging issue.

Distributed resource allocation in can benefit from being deployed at the edge of network, reducing delay and energy consumption. However, distribution optimization in a parallel manner for performance improvement brings to light several challenges. How could we design an efficient parallel optimization mechanism for large-scale task scheduling? How could we guarantee convergence to a resource allocation scheme and performance optimality given task management problem?

To answer these questions, we present an optimal task allocation mechanism for three-tier model based edge computing system, which can make full use of resource among MT, small cell base station (SBS), and macro cell base station (MBS). The proposed mechanism can dispatch task in resource allocation among different MTs, SBS, and MBS,

This work was supported by the National Natural Science Foundation of China (Grant Nos. 61802221, 61802220, 61671165), and the Natural Science Foundation of Guangxi Province under grant 2017GXNSFAA198192, the Innovation Project of Guangxi Graduate Education under grant YCSW2019141, and the Key-Area Research and Development Program for Guangdong Province 2019B010136001, the Peng Cheng Laboratory Project of Guangdong Province PCL2018KP005 and PCL2018KP004. (*Corresponding authors*: Xinghan Wang, Tingting Yang).

Xiaoxiong Zhong and Xinghan Wang are with Peng Cheng Laboratory, Shenzhen 518000, P. R. China; Xiaoxiong Zhong is also with the Graduate School at Shenzhen, Tsinghua University, Shenzhen 518055, P. R. China. (email: {xixzhong, csxhwang}@gmail.com.

Yuanyuan Yang is with Department of Electrical and Computer Engineering, Stony Brook University, Stony Brook, NY11794, USA. (e-mail: yuanyuan.yang@stonybrook.edu).

Yang Qin is with the Department of Computer Science and Technology, Harbin Institute of Technology (Shenzhen), Shenzhen, 518055, China. (e-mail: csyqin@hit.edu.cn).

Xiaoke Ma is School of Computer Science and Technology, Xidian University, Xi'an, Shaanxi, 710071, P. R. China. (email: xkma@xidian.edu.cn)

Tingting Yang is with the School of Electrical Engineering and Intelligentization, Dongguan University of Technology, Dongguan, 523000, P. R. China. (email: yangtingting820523@163.com).

featuring distributed optimization and efficient task allocation mechanism, to reach intelligent and efficient resource management in MEC. The contributions of this article are as follows:

In this paper, we exploit a three-tier model for task management, which can make full use of resource among MTs (local terminals, LTs), SBS and MBS, reducing delay and energy.

1) It is more difficult to guarantee convergence in multi-block ADMM, only if its optimal function is linear. In the proposed optimal scheme, we exploit regulation linear method to make the optimal function to be linear, guaranteeing its convergence and it can run in a parallel manner and its global and local variables can update simultaneously, which can reduce computational complexity for achieving good scalability.
2) Our joint optimization problem is a nonlinear 0-1 integer programming problem, which can be solved using Logarithmic Smoothing by modifying the inequality constrain as equality constrain in decision variables and smoothing function. Also, we prove the convergence of the proposed Logarithmic Smoothing based optimization scheme.
3) In order to obtain the optimal values of the global variables, we use Newton method to calculate the partial derivative of the global variables. For the purpose of accelerating the convergence of the global variables, we adopt the conjugate gradient to solve it. After obtaining the partial derivative of the global variables, we use linear search technique to get the optimal step and update the global variables.
4) In order to solve the sub-optimal problem, in which the objective function is convex with more variables updating simultaneously, we adopt the cyclic block coordinate gradient projection (CBGP) method to address the sub-optimal problem, which can guarantee convergence and reduce computational complexity.
5) We conduct extensive experiments to evaluate the performance of the proposed mechanism. With the numerical results, we show that the proposed method can guarantee the convergence and effectively reduce delay and energy consumption.

The remainder of this paper is organized as follows. We review the related work in Section II. In Section III, we describe the system model, energy model and computation model. We propose an improved parallel multi-block ADMM based task allocation framework for MEC in Section IV. We evaluate the performance of the proposed mechanism and provide illustrative results in Section V and conclude the paper in Section VI.

## II. RELATED WORK

Recently, some optimization algorithm based resource management schemes have been proposed for MEC [4-17]. Chen *et al*. [4] developed an optimization framework for software defined ultra-dense networks, minimizing the delay and saving the energy, which includes two sub-optimization problems: task placement and resource allocation. Consider cooperation behaviors, Xu *et al*. [5] proposed an online algorithm for service caching and task offloading in MEC systems, which jointly considers service heterogeneity, system dynamics, spatial demand and distributed coordination. Similarly, for heterogeneous services, Tan *et al*. [6] formulated a virtual resource allocation problem, and proposed ADMM algorithm to solve it, with jointly considering user association, power control and resources allocation. Zhou *et al*. [7] formulated the virtual resource allocation strategy as a joint optimization problem and used ADMM to address it, considering virtualization, caching and computing. Considering edge nodes and mobile users with time-dependent requests, Zheng *et al*. [8] proposed a convergent and scalable Stackelberg game for edge caching, and used a Stackelberg game based ADMM to solve storage allocation game and user allocation game in a distributed manner. Wang *et al*. [9] presented a parallel optimization framework which leverages the vertical cooperation among devices, edge nodes and cloud servers, and the horizontal cooperation between edge nodes by ADMM method to address the resource allocation optimal problem. Zhou *et al*. [10] studied the energy-efficient workload offloading problem and propose a low-complexity distributed solution based on consensus alternating direction method of multipliers (ADMM) for vehicular networks edge computing service provisioning. In [11], Dai *et al*. proposed a novel two-tier computation offloading framework in heterogeneous networks, which formulated joint computation offloading and user association problem for multi-task MEC system to minimize overall energy consumption through computation resource allocation and transmission power allocation. Wang *et al*. [12] proposed a heterogeneous multi-layer MEC framework, where data that cannot be timely processed at the edge are allowed to be offloaded to the upper layer MEC servers and the cloud center. Their goal is minimizing the system latency, i.e., the total computing and transmission time on all layers for the data generated by the edge devices, by jointly coordinating the task assignment, computing, and transmission resources in the framework. Li *et al*. [13] studied collaborative cache allocation and task scheduling in edge computing and modeled the task scheduling problem as a weighted bipartite graph, which can reduce latency. In [14], Alameddine *et al*. mathematically formulated the dynamic task offloading and scheduling as a mixed integer program and exploited a logic based bender decomposition approach to efficiently address the problem to optimality for low latency IoT service in MEC. For the ultra-reliable low-latency requirements in mission-critical applications, Liu *et al*. [15] proposed a new system design, where probabilistic and statistical constraints are imposed on task queue lengths, by applying extreme value theory and marrying tools from Lyapunov optimization and matching theory for user-server association, and dynamic task offloading and resource allocation. Meng *et al*. [16] studied the online deadline-aware task dispatching and scheduling in edge computing, which jointly considers with bandwidth constraint using joint optimization of networking and computing resource to meet the deadlines and proposed an online algorithm Dedas, which greedily schedules newly arriving tasks and considers whether to replace some existing tasks in order to make the new deadlines satisfied. In [17], Kim *et al*. proposed dual-side control algorithms for cost-delay tradeoff in MEC, including two optimization problems where the objective is to minimize costs subject to queue stability under two scenarios: a competition scenario and a cooperation scenario.

Most of the existing resource allocation algorithms in MEC were proposed based on the environment without large-scale model and they perform well only when the assumptions do hold. However, the resource allocation algorithms in real-world large-scale MEC are often too complex to be designed by traditional cloud computing. Till now, there is still little attention being paid on applying optimization algorithms for resource management of large-scale MEC in a parallel manner, which will cause a poor system performance. Hence, how to design an efficient parallel task allocation scheme for large-scale MEC is a challenging issue. This paper aims to propose a solution to address this problem.

III. MODEL FOR TASK ALLOCATION MECHANISM

In this section, we will describe the system model, computation model and energy consumption model for the proposed task allocation mechanism in MEC.

1. System model

In this paper, we consider a large scale mobile edge computing system, which is composed of many groups, and we denote the set of groups as $GR = \{gr_1, gr_2, ..., gr_n\}$, each group is made up of a bunch of BS, including an MBS and some SBS, denoted as $BS = \{b_1, b_2, ..., b_i\}$, where $b_1$ represent an MBS and each element of set $\{b_2, ..., b_i\}$ represents a SBS, each BS is equipped with MEC-server.

Each task that enters the coverage of group will be processed in local or in there-tier computation model or in MBS. Each task can be separated into subtasks. The set of tasks is $H = \{h_1, h_2, ..., h_j\}$, each task $h_j$ can be expressed through a vector $h_j = \{c_j, t_j^{max}, u_j\}$, $c_j$ denotes the data size of the task, $t_j^{max}$ is the deadline of $h_j$, and $u_j$ is the number of CPU cycle for computing one bit of task $h_j$.

In order to make task algorithm suitable for the realistic large-scale environment, and leverage a large number of the computation server resources, we adopt a parallel approach of local computation, MBS computation and there tier computation, which jointly considers path in resource allocation mechanism, as shown in Fig. 1. The path 1{task3, Local3, Virtual destination} is an example for local computing model, denoted by the blue line. The path 2 {task1, Local1, AP_1, SBS1, FU_2, MBS, Virtual destination}, is an example for three-tier computing model, denoted by the red line (also, the paths {task2, AP_2, SBS2, FU_3, MBS, Virtual destination}, {task1, Local1, AP_1, FU_1, MBS, Virtual destination} and {task3, Local3, AP_3, SBS3, Virtual destination} are examples for three-tier computing model). The path 3 {task1, SBS1, Virtual destination} is an example for SBS computing model, denoted by the green line. The path 4 {task2, MBS, Virtual destination} is an example for MBS computing model, denoted by the purple line. If an adjacent unit help local task transmit data, we call adjacent unit the parent of local task. By extension, the path can be defined as $Path_{hj \rightarrow SBS/MBS/Local}$ = {(task, link$^1$, forwarding$^1$($h_j$)), (forwarding$^1$($h_j$), link$^2$, forwarding$^2$($h_j$))... (forwarding$^{n-1}$($h_j$), link$^{n-1}$, SBS/MBS/Local)}.

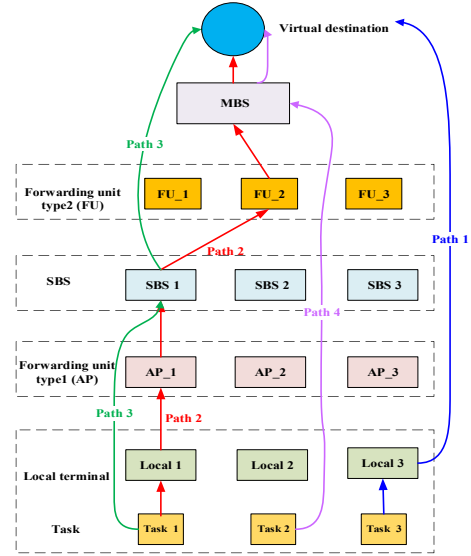

Fig. 1. Three- tier framework for MEC task scheduling mechanism.

The large-scale edge computing network can be described as a graph $G = \{v, \varepsilon\}$, where $v$ and $\varepsilon$ denote vertices and edges, respectively. The vertices are computing units and forwarding units, in which computing units (like MEC server deployed on SBS or MBS) provide computing services, and forwarding units (like access point and switch) provide directing function. A virtual destination is introduced in order to let all the tasks flow to the same destination. Virtual destination is a sign of the end of tasks processing. Each vertex is associated with a source and a sink pair $\{v_{source}, v_{destination}\}$. We use $f_v$ to denote a path that traversing the vertice $v$, and let $F = \bigcup f_v$ to denote the feasible paths from local terminal to processing node in the large-scale edge computing. A path $f \in F$ consists of a sequence of links, a series forwarding units, a source and a processing node. We use the $Y_{h_j} = \{y_{h_j, f}\} f \in F$ represents possible paths of the part of task $h_j$ indexed by $F$.

So we can conclude the aggregation load of each forwarding unit (i.e., $load_{forwarding}$) and each link $load_{link}$.

$$load_{forwarding} = \sum_{h_j \in H} \sum_{forwarding \in f} y_{h_j, f} c_j \quad \forall forwarding \in v \quad (1)$$

and

$$load_{link} = \sum_{h_j \in H} \sum_{link \in f} y_{h_j,f} c_j \quad \forall link \in \varepsilon \qquad (2)$$

We adopt a linear latency model since buffer utilization and packet loss increases as the bitrate grows [18]. Hence, we can get the latency of the forwarding unit $Delay_{forwarding}$ and the overall transmission latency on wireless channels $Delay_{link}$, which can be expressed as:

$$Delay_{forwarding} = (o_1 load_{forwarding} + o_2) \times load_{forwarding} \qquad (3)$$

and

$$Delay_{link} = \frac{load_{link}}{capacity_{link}} \qquad (4)$$

where $o_1$ and $o_2$ are parameters for linear model, which are related to forwarding unit, $capacity_{link}$ is the capacity of link.

### 2. Local computing model & energy consumption model

First, the local process time should be defined according to the following equation:

$$T_{local}^{(h_j)} = z_j \frac{c_j u_j}{f_{local}} \qquad (5)$$

For local task computing, we define the local energy consumption model:

$$E^{(h_j)} = c_j u_j e_{local} \qquad (6)$$

where $f_{local}$ is the process capability of local devices, $z_j$ is a binary variable and has $z_j \in \{0,1\}$, it has $z_j=1$ if the task $h_j$ decides to process locally; it has $z_j=0$ if the task is not processed locally and $e_{local}$ is the energy consumption per CPU cycle of local device.

### 3. MBS computation model

In this paper, we describe the computation model from a delay perspective. The overall delay in MBS $b_1$ can be expressed as:

$$T^{(b_1,h_j)} = T_{up}^{(b_1,h_j)} + T_{execute}^{(b_1,h_j)} \qquad (7)$$

where $T_{up}^{(b_1,h_j)}$ is the uploading time, and $T_{execute}^{(b_1,h_j)}$ is the calculation time.

$T_{up}^{(b_1,h_j)}$ can be defined as:

$$T_{up}^{(b_1,h_j)} = y_{1j} \frac{c_j}{B_{MBS} \log_2 \left(1 + \frac{p^{(b_1,h_j)} G^{(b_1,h_j)}}{\sigma^2}\right)} \qquad (8)$$

where $p^{(b_1,h_j)}$ is the transmission power density of the MBS $b_1$, $G^{(b_1,h_j)}$ is the channel gain between the local terminal and the MBS, $\sigma^2$ is the power of the additive white Gaussian noise, $y_{1j}$ is a binary variable and has $y_{1j} \in \{0,1\}$, it has $y_{1j}=1$ if the task $h_j$ decides to upload to MBS; it has $y_{1j}=0$ if the task is not processed by the MBS.

The calculation time in MEC server deployed on MBS $T_{execute}^{(b_1,h_j)}$ can be expressed:

$$T_{execute}^{(b_1,h_j)} = y_{1j} \frac{c_j u_j}{f^{(b_1)}} \qquad (9)$$

where $f^{(b_1)}$ is the maximum number of task that MEC-Server can handle.

### 4. Three-tier computation model

If the MEC-server deployed on SBSs don't have sufficient computation resources, it will uploaded remaining part of its to MEC-server deployed on MBS, which can effectively reduce time delay and energy consumption, otherwise, it will solve problem independently. So we divided the task into three parts $c_{0j}$, $c_{1j}$ and $c_{ij}$, $c_{0j}$ represents the part of subtask is processed in local terminal, $c_{1j}$ represents the part of subtask is processed in MBS $b_1$, and $c_{ij}$ represents the remaining part of task is processed in SBS $b_i$, we can conclude that $c_{0j} + c_{1j} + c_{ij} = c_j$. Therefore, the over delay in SBS $b_i$ can be expressed as:

$$T^{(b_i,h_j)} = x_{ij} \frac{c_{0j} u_j}{f_{local}} + x_{ij} \frac{c_j - c_{0j}}{B_{SBS} \log_2 \left(1 + \frac{p^{(b_i,h_j)} G^{(b_i,h_j)}}{\sigma^2 + \sum_{i'=2,i' \neq i}^{|BS|} \sum_{j'=1,j' \neq j}^{|H|} p^{(b_{i'},h_{j'})} G^{(b_{i'},h_{j'})}}\right)}$$

$$+ x_{ij} \left( \sum_{\substack{link \in Path_{h_j \to b_i} \\ forwarding \in Path_{h_j \to b_i}}} Delay_{link} + Delay_{forwarding} \right) + x_{ij} \frac{c_{ij} u_j}{h^{(b_i,h_j)} f^{(b_i)}} + x_{ij} \frac{c_{1j} u_j}{f^{(b_1)}}$$

(10)

where the first item indicates transmission time from local terminal to SBS $b_i$, the second item indicates computation time of its subtasks, the third item indicates offloading time on wired line from SBS $b_i$ to MBS $b_1$, and the forth part show that computation time which processed by MBS $b_1$. $h^{(b_i,h_j)}$ is a binary variable and has $h^{(b_i,h_j)} \in [0,1]$, which stands for the computation resource allocation about MEC server deployed on the SBS or not.

### 5. Edge energy consumption model

In our model, we exploit the following energy consumption model [19]:

$$E^{(h_1,h_j)} = y_{1j}\left(E_{upload}^{(h_1,h_j)} + E_{execute}^{(h_1,h_j)}\right)$$
$$= y_{1j}\left(P_U T_{up}^{(h_1,h_j)} + c_j u_j e_{MBS}\right) \quad (11)$$

where $P_U$ is the transmit power, $e_1$ is the energy consumption per CPU cycle of MBS.

$$E^{(b_i,h_j)} = x_{ij}\left(E_{subtask\_local} + E_{upload}^{(b_i,h_j)} + E_{execute}^{(b_i,h_j)} + E_{transfer}^{(b_i,b_1,c_{1j})} + E_{execute}^{(b_1,c_{1j})}\right)$$
$$= x_{ij}\left(c_{0j}u_j e_{local} + P_U T_{upload}^{(b_i,h_j)} + c_j u_j e_{SBS} + P_{SBS\to MBS}T_{transfer}^{(b_i,b_1,c_{1j})} + c_{1j}u_j e_{MBS}\right) \quad (12)$$

where $P_{SBS\to MBS}$ is offloading power between SBS and MBS, $P_{SBS}$ is the SBS computation power consumption(in watt),and $e_{SBS}$ is the energy consumption per CPU cycle of SBS. The first item indicates the transmission energy consumption from local to SBS, the second item indicates the computation energy consumption of subtask $c_{ij}$ in SBS, the third item is the transmission energy consumption of $c_{1j}$ from SBS to MBS, the forth item presents us with the computation energy consumption of remaining task $c_{1j}$ in MBS.

## IV. PARALLEL MULTI-BLOCK ADMM BASED TASK ALLOCATION MECHANISM

In this section, based on the above models, we will describe the improved multi-block ADMM based optimal task allocation mechanism in detail, which jointly considers delay and energy in optimization utility, as shown in **P1**.

The (13-1) guarantees that the time of the task that uploaded into MEC Server or the time of the task that processed in local terminals cannot exceed the deadline time. The constraint (13-3) guarantees that the sum of computation resource allocated to all offloading task cannot exceed the total amount of computation resource of MEC server deployed on SBS. The constraint (13-4) indicates each task is either executed locally, or by one of MEC server which deployed either on the MBS or on the SBS.

It has $|H|\times|BS|$ variables in our optimization problem, where $|BS|$ is the number of BSs, $|H|$ is the number of the tasks. Thus, the computational complexity significantly increases as the number of tasks and BSs. To solve the large-scale optimization problem, we propose a scalable and practical distributed method based on multi-block ADMM with using Logarithmic Smoothing [20] by modifying the inequality constrain as equality constrains in decision variables and smoothing function and using the cyclic block coordinate gradient projection (CBGP) method [21] to address the sub-optimal problem.

In order to make the problem separable, we develop parallel optimization based ADMM and introduce the local copies of the global variables and the local copies of the variables $x_{ij}$, $z_j$, $y_{1j}$, which are defined as $\hat{x}_{ij}$, $\hat{z}_j$, $\hat{y}_{1j}$.

$$\mathbf{P1}: \min \alpha\left(\sum_{j=1}^{|H|}z_j T_{local}^{(h_j)} + \sum_{i=2}^{|BS|}\sum_{j=1}^{|H|}x_{ij}T^{(b_i,h_j)} + \sum_{j=1}^{|H|}y_{1j}T^{(b_1,h_j)}\right) + (1-\alpha)\left(\sum_{j=1}^{|H|}z_j E^{(h_j)} + \sum_{i=2}^{|BS|}\sum_{j=1}^{|H|}x_{ij}E^{(b_i,h_j)} + \sum_{j=1}^{|H|}y_{1j}E^{(b_1,h_j)}\right)$$

$$s.t.\begin{cases} z_j T_{local}^{(h_j)} + \sum_{i=2}^{|BS|}x_{ij}T^{(b_i,h_j)} + y_{1j}T^{(b_1,h_j)} \leq t_j^{\max} & (13\text{-}1) \\ x_{ij}\in\{0,1\}, y_{1j}\in\{0,1\}, z_j\in\{0,1\} & (13\text{-}2) \\ \sum_{j=1}^{|BS|}h^{(b_i,h_j)}x_{ij}\leq 1, 0 < h^{(b_i,h_j)}\leq 1 & (13\text{-}3) \\ \sum_{i=2}^{|BS|}x_{ij} + z_j + y_{1j} = 1 & (13\text{-}4) \\ 0\leq \alpha \leq 1 & (13\text{-}5) \\ c_{1j} + c_{ij} + c_{0j} = c_j, 0\leq c_{0j}\leq c_j, 0\leq c_{ij}\leq c_j, 0\leq c_{1j}\leq c_j & (13\text{-}6) \\ i\in\{2,3...,|BS|\}, j\in\{1,2,3...,|H|\} & (13\text{-}7) \end{cases} \quad (13)$$

Then we can obtain the equivalent global consensus version as **P2**.

In order to express more easily, we define the local variable set with some constrains. The term $x_{ij}\dfrac{1}{h^{(b_i,h_j)}}$ in contains of the objective function **P1** will lead to be non-convex. To deal with these sub-problems, we set $\dfrac{1}{h^{(b_i,h_j)}}$ as $r^{(b_i,h_j)}$ we define auxiliary variables as follows $R_{ij} = x_{ij}r^{(b_i,h_j)}$.

We add some constrains to above equation in order to overcome the difficulty of the problem intractable. We use the reformulation linearization technology to replace above equation.

$$\mathbf{P2}: \ \min \alpha \left( \sum_{j=1}^{|H|} \hat{z}_j T_{local}^{(h_j)} + \sum_{i=2}^{|BS|}\sum_{j=1}^{|H|} \hat{x}_{ij} T^{(b_i,h_j)} + \sum_{j=1}^{|H|} \hat{y}_{1j} T^{(b_1,h_j)} \right) + (1-\alpha)\left( \sum_{j=1}^{|H|} \hat{z}_j E^{(h_j)} + \sum_{i=2}^{|BS|}\sum_{j=1}^{|H|} \hat{x}_{ij} E^{(b_i,h_j)} + \sum_{j=1}^{|H|} \hat{y}_{1j} E^{(b_1,h_j)} \right)$$

$$s.t \begin{cases} z_j T_{local}^{(h_j)} + \sum_{i=2}^{|BS|} x_{ij} T^{(b_i,h_j)} + y_{1j} T^{(b_1,h_j)} \le t_j^{\max} \\ \hat{x}_{ij} \in \{0,1\}, \ \hat{y}_{1j} \in \{0,1\}, \ \hat{z}_j \in \{0,1\} \\ \sum_{j=1}^{|BS|} h^{(b_i,h_j)} \hat{x}_{ij} \le 1 \\ 0 < h^{(b_i,h_j)} \le 1 \\ \sum_{i=2}^{|BS|} x_{ij} + z_j + y_{1j} = 1 \\ x_{ij} = \hat{x}_{ij}, \ z_j = \hat{z}_j, \ y_{1j} = \hat{y}_{1j} \\ c_{1j} + c_{ij} + c_{0j} = c_j, \ 0 \le c_{0j} \le c_j, \ 0 \le c_{ij} \le c_j, \ 0 \le c_{1j} \le c_j \\ 0 \le \alpha \le 1, \ i \in \{2,3...,|BS|\}, j \in \{1,2,3...,|H|\} \end{cases} \quad (14)$$

**Theorem 1**: The objective function in **P2** can be changed as a linear function.

**Proof**: Please refer to the **Appendix A**.

Therefore, it is easy to divide the objective function into independent terms. As a result, **P2** can be decomposed equivalently into the following sub-problems (**P3**, **P4** and **P5**).

A. Augmented Lagrangian and ADMM sequential iterations

The local variables can be expressed as **P3**, **P4** and **P5**.

$$\mathbf{P3}: \min \alpha \sum_{i=2}^{|BS|}\sum_{j=1}^{|H|} \hat{x}_{ij} T^{(b_i,h_j)} + (1-\alpha)\sum_{i=2}^{|BS|}\sum_{j=1}^{|H|} \hat{x}_{ij} E^{(b_i,h_j)} + \sum_{i=2}^{|BS|}\sum_{j=1}^{|H|} \frac{\rho}{2}(\hat{x}_{ij} - x_{ij})^2$$

$$s.t \begin{cases} \hat{x}_{ij} \in \{0,1\} \\ \sum_{j=1}^{|H|} \frac{1}{r^{(b_i,h_j)}} \le 1 \\ \hat{x}_{ij} \le R_{ij} \le \hat{x}_{ij} \frac{1}{h_{\min}^{(b_i,h_j)}} \\ R_{ij} \le r^{(b_i,h_j)} + \hat{x}_{ij} - 1 \\ R_{ij} \ge r^{(b_i,h_j)} + \hat{x}_{ij} \frac{1}{h_{\min}^{(b_i,h_j)}} - \frac{1}{h_{\min}^{(b_i,h_j)}} \\ c_{1j} + c_{ij} + c_{0j} = c_j, \ 0 \le c_{0j} \le c_j, \ 0 \le c_{ij} \le c_j, \ 0 \le c_{1j} \le c_j \\ i \in \{2,3...,|BS|\}, j \in \{1,2,3...,|H|\} \end{cases} \quad (15)$$

$$\mathbf{P4}: \min \alpha \sum_{j=1}^{|H|} \hat{z}_j T_{local}^{(h_j)} + (1-\alpha)\sum_{j=1}^{|H|} \hat{z}_j E^{(h_j)} + \sum_{j=1}^{|H|} \frac{\rho}{2}\left(\hat{z}_j - z_j\right)^2$$
$$s.t \quad \hat{z}_j \in \{0,1\} \quad (16)$$

$$\mathbf{P5}: \min \alpha \sum_{j=1}^{|H|} \hat{y}_{1j} T^{(b_1,h_j)} + (1-\alpha)\sum_{j=1}^{|H|} \hat{y}_{1j} E^{(b_1,h_j)} + \sum_{j=1}^{|H|} \frac{\rho}{2}\left(\hat{y}_{1j} - y_{1j}\right)^2$$
$$s.t \quad \hat{y}_{1j} \in \{0,1\} \quad (17)$$

The global variables can be expressed as **P6**.

Hence, the Lagrange multipliers can be expressed by (19).

$$\begin{cases} \alpha_{ij}^{k+1} = \alpha_{ij}^k + \rho\left(\hat{x}_{ij}^{k+1} - x_{ij}^{k+1}\right) \\ \beta_j^{k+1} = \beta_j^k + \rho\left(\hat{z}_j^{k+1} - z_j^{k+1}\right) \\ \gamma_j^{k+1} = \gamma_j^k + \rho\left(\hat{y}_{1j}^{k+1} - y_{1j}^{k+1}\right) \\ i \in \{2,3...,|BS|\}, j \in \{1,2,3...,|H|\} \end{cases} \quad (19)$$

B. Local Variables Update

Obviously, the binary variable is non-convex, in order to transform non-convex set into convex set, we need to relax binary variable $\hat{x}_{ij}$ with a real value such that $\sum_{i=1}^{|BS|}\sum_{j=1}^{|H|} \hat{x}_{ij} - \hat{x}_{ij}^2 \le 0$ and $0 \le \hat{x}_{ij} \le 1$.

**Theorem 2**: **P3** can be changed as **P7**.

**Proof**: Please refer to the **Appendix B**.

$$P6: \min \sum_{i=2}^{|BS|}\sum_{j=1}^{|H|}\alpha_{ij}(\hat{x}_{ij}^{k+1}-x_{ij})+\sum_{i=2}^{|BS|}\sum_{j=1}^{|H|}\frac{\rho}{2}(\hat{x}_{ij}^{k+1}-x_{ij})^2+\sum_{j=1}^{|H|}\beta_j\left(\hat{z}_j^{k+1}-z_j\right)+\sum_{j=1}^{|H|}\frac{\rho}{2}\left(\hat{z}_j^{k+1}-z_j\right)^2+\sum_{j=1}^{|H|}\gamma_j\left(\hat{y}_{1j}-y_{1j}\right)+\sum_{j=1}^{|H|}\frac{\rho}{2}\left(\hat{y}_{1j}-y_{1j}\right)^2$$

$$s.t \begin{cases} z_j T_{local}^{(h_j)}+\sum_{i=2}^{|BS|}x_{ij}T^{(b_i,h_j)}+y_{1j}T^{(b_1,h_j)}\leq t_j^{\max} \\ \sum_{i=2}^{|BS|}x_{ij}+z_j+y_{1j}=1 \\ x_{ij}\in\{0,1\},\ y_{1j}\in\{0,1\},\ z_j\in\{0,1\} \\ i\in\{2,3...,|BS|\},\ j\in\{1,2,3...,|H|\} \end{cases}$$

(18)

$$P7: \min Q(\hat{x}_{ij},R_{ij},c_{1j},c_{ij},c_{0j})=\alpha\sum_{i=2}^{|BS|}\sum_{j=1}^{|H|}\hat{x}_{ij}T^{(b_i,h_j)}+(1-\alpha)\sum_{i=2}^{|BS|}\sum_{j=1}^{|H|}\hat{x}_{ij}E^{(b_i,h_j)}+\sum_{i=2}^{|BS|}\sum_{j=1}^{|H|}\frac{\rho}{2}(\hat{x}_{ij}-x_{ij})^2+\delta_{ij}\left(x_{ij}^{k+1}-\left(x_{ij}^k\right)^2-\left\langle\nabla\left(x_{ij}^k\right)^2,x_{ij}^{k+1}-x_{ij}^k\right\rangle\right)$$

$$s.t \begin{cases} 0\leq\hat{x}_{ij}\leq 1,\ \sum_{j=1}^{|H|}\frac{1}{r^{(b_i,h_j)}}\leq 1 \\ \hat{x}_{ij}\leq R_{ij}\leq\hat{x}_{ij}\frac{1}{h_{\min}^{(b_i,h_j)}} \\ R_{ij}\leq r^{(b_i,h_j)}+\hat{x}_{ij}-1,\ R_{ij}\geq r^{(b_i,h_j)}+\hat{x}_{ij}\frac{1}{h_{\min}^{(b_i,h_j)}}-\frac{1}{h_{\min}^{(b_i,h_j)}} \\ c_{1j}+c_{ij}+c_{0j}=c_j,\ 0\leq c_{0j}\leq c_j,\ 0\leq c_{ij}\leq c_j,\ 0\leq c_{1j}\leq c_j \\ i\in\{2,3...,|BS|\},\ j\in\{1,2,3...,|H|\} \end{cases}$$

(20)

According to the observations in **P7**, the objective function is convex, therefore we have **Theorem 3**.

**Theorem 3**: The **P7** can be solved by CBGP method.

**Proof**: Please refer to the **Appendix C**.

According to **Theorem 3**, we solve the problem by fixing four of five variables and deriving the remaining one, as shown in **Algorithm 1**.

---

**Algorithm 1** Variables update in CBGP

**Input:**
The amount of computing resource are required for the task : $c_{ij}$,
Parameter setting: $\alpha\in(0,1)$
The rounds of iterations: $\kappa$
**Output**: $x_{ij}, c_{0j}, c_{ij}, c_{1j}, R_{ij}$

1: Initialize $\hat{x}_{ij},\varepsilon_{ij},\phi_{ij},\varpi_{ij},\theta_{ij},\vartheta_{ij},\sigma_{ij},c_{0j},c_{ij},c_{1j},R_{ij}$
2: **while** $k<\kappa$ **do**
3:   $R_{ij}^{(k+1)}=P_{\Omega_{R_{ij}}}(R_{ij}^{(k)}-\varepsilon_{ij}\nabla Q(R_{ij}^{(k)}))$
4:   $c_{0j}^{(k+1)}=P_{\Omega_{c_{0j}}}(c_{0j}^{(k)}-\chi_j\nabla Q(c_{0j}^{(k)}))$
5:   $c_{1j}^{(k+1)}=P_{\Omega_{c_{1j}}}(c_{1j}^{(k)}-\kappa_j\nabla Q(c_{1j}^{(k)}))$
6:   $c_{ij}=c_j-c_{0j}-c_{1j}$
7:   $x_{ij}^*\leftarrow$ solving the problem with fixed $c_{0j},c_{ij},c_{1j},R_{ij}$
8:   $k++$
9: **end while**

---

The objective function of **P4** is linear and strongly convex, which can be solved by comparing the two possible solutions. So the solution of **P4** can be solved as following:

$$\begin{cases} \hat{z}_j=0\ if\ \alpha T_{local}^{(h_j)}+(1-\alpha)E^{(h_j)}+\frac{\rho}{2}(1-z_j)>0 \\ \hat{z}_j=1\ if\ \alpha T_{local}^{(h_j)}+(1-\alpha)E^{(h_j)}+\frac{\rho}{2}(1-z_j)<0 \end{cases}$$ (21)

Similarly, the objective function of **P5** is linear and strongly convex, which can be solved by comparing the two possible solutions. So the solution of **P5** can be solved as following:

$$\begin{cases} \hat{y}_{1j}=0\ if\ \alpha T^{(b_1,h_j)}+(1-\alpha)E^{(b_1,h_j)}+\frac{\rho}{2}(1-y_{1j})>0 \\ \hat{y}_{1j}=1\ if\ \alpha T^{(b_1,h_j)}+(1-\alpha)E^{(b_1,h_j)}+\frac{\rho}{2}(1-y_{1j})<0 \end{cases}$$ (22)

In the sub-problem **P6**, the objective function is nonlinear due to the nonlinear integer value $x_{ij}$, $y_{1j}$ and $z_j$, the similar method can be adopted in solving the integer constrain problems. Here, we use Logarithmic Smoothing to solve the nonlinear integer problem, the smoothing function aims to eliminate all the integer constrains. Secondly, in order to make the value close to zero or one, the objective function must be added to some penalty functions. So the objective function can be described as **P8**.

Next we use Newton's method to solve the problems. Firstly, we add some auxiliary variables to transform the inequality constraints to the equality constraints, (24) and (25).

$$z_j T_{local}^{(h_j)} + \sum_{i=2}^{|BS|} x_{ij} T^{(b_i,h_j)} + y_{1j} T^{(b_1,h_j)} + m_j = t_j^{max} \quad (24)$$

$$\begin{cases} \nabla_x G(x,y,z,m) + T^{(b_i,h_j)} \upsilon_j + \varsigma_j = 0 \\ \nabla_y G(x,y,z,m) + T^{(b_1,h_j)} \upsilon_j + \varsigma_j = 0 \\ \nabla_z G(x,y,z,m) + T_{local}^{(h_j)} \upsilon_j + \varsigma_j = 0 \\ \nabla_m G(x,y,z,m) + \upsilon_j = 0 \\ z_j T_{local}^{(h_j)} + \sum_{i=2}^{|BS|} x_{ij} T^{(b_i,h_j)} + y_{1j} T^{(b_1,h_j)} + m_j - t_j^{max} = 0 \\ \sum_{i=2}^{|BS|} x_{ij} + z_j + y_{1j} - 1 = 0 \end{cases} \quad (25)$$

where $m_j \geq 0$.

Then, the **P8** can be changed as **P9**. Hence, the Lagrangian equation is shown in (27). From (27), we can obtain the first order partial derivative about $x, y, z, m, \upsilon, \varsigma$.

**Theorem 4**: **P9** can be solved by the combination of conjugate gradient, Newton and linear search techniques based algorithm with Logarithmic Smoothing.

**Proof**: Applying Newton's method directly, we can obtain the following equation:

$$\begin{bmatrix} g_1 & g_2 & g_3 & g_4 & T^{(b_i,h_j)} & 1 \\ g_5 & g_6 & g_7 & g_8 & T^{(b_1,h_j)} & 1 \\ g_9 & g_{10} & g_{11} & g_{12} & T_{local}^{(h_j)} & 1 \\ g_{13} & g_{14} & g_{15} & g_{16} & 1 & 0 \\ T^{(b_i,h_j)} & T^{(b_1,h_j)} & T_{local}^{(h_j)} & 1 & 0 & 0 \\ 1 & 1 & 1 & 0 & 0 & 0 \end{bmatrix} \begin{bmatrix} \Delta x \\ \Delta y \\ \Delta z \\ \Delta m \\ \Delta \upsilon \\ \Delta \varsigma \end{bmatrix} = \begin{bmatrix} h_1 \\ h_2 \\ h_3 \\ h_4 \\ h_5 \\ h_6 \end{bmatrix} \quad (28)$$

where

$$\begin{cases} \Delta x = [\Delta x_{2j}, \Delta x_{3j}, ..., \Delta x_{ij}] \quad j=1,2,...|H| \\ \Delta y = [\Delta y_{11}, \Delta y_{12}, ..., \Delta y_{1|H|}] \\ \Delta Z = [\Delta Z_1, \Delta Z_2, ..., \Delta Z_{|H|}] \end{cases} \quad (29)$$

$$\begin{cases} g_1 = \nabla_{xx} G(x,y,z,m) & g_2 = \nabla_{xy} G(x,y,z,m) & g_3 = \nabla_{xz} G(x,y,z,m) \\ g_4 = \nabla_{xm} G(x,y,z,m) & g_5 = \nabla_{yx} G(x,y,z,m) & g_6 = \nabla_{yy} G(x,y,z,m) \\ g_7 = \nabla_{yz} G(x,y,z,m) & g_8 = \nabla_{ym} G(x,y,z,m) & g_9 = \nabla_{zx} G(x,y,z,m) \\ g_{10} = \nabla_{zy} G(x,y,z,m) & g_{11} = \nabla_{zz} G(x,y,z,m) & g_{12} = \nabla_{zm} G(x,y,z,m) \\ g_{13} = \nabla_{mx} G(x,y,z,m) & g_{14} = \nabla_{my} G(x,y,z,m) & g_{15} = \nabla_{mz} G(x,y,z,m) \\ g_{16} = \nabla_{mm} G(x,y,z,m) \end{cases} \quad (30)$$

and

$$\begin{cases} h_1 = -\left(\nabla_x G(x,y,z,m) + T^{(b_i,h_j)} \upsilon_j + \varsigma_j\right) \\ h_2 = -\left(\nabla_y G(x,y,z,m) + T^{(b_1,h_j)} \upsilon_j + \varsigma_j\right) \\ h_3 = -\left(\nabla_z G(x,y,z,m) + T_{local}^{(h_j)} \upsilon_j + \varsigma_j\right) \\ h_4 = -\left(\nabla_m G(x,y,z,m) + \upsilon_j\right) \\ h_5 = -\left(z_j T_{local}^{(h_j)} + \sum_{i=2}^{|BS|} x_{ij} T^{(b_i,h_j)} + y_{1j} T^{(b_1,h_j)} + m_j - t_j^{max}\right) \\ h_6 = -\left(\sum_{i=2}^{|BS|} x_{ij} + z_j + y_{1j} - 1\right) \end{cases} \quad (31)$$

We observe that the (40) can be reduced as follow:

$$\begin{bmatrix} H & A_2 \\ A_1 & 0 \end{bmatrix} \begin{bmatrix} U \\ \Delta \varsigma \end{bmatrix} = \begin{bmatrix} M \\ h_6 \end{bmatrix} \quad (32)$$

where

$$U = [\Delta x, \Delta y, \Delta z]^T, \quad H = Diag(D_1, D_2, D_3)$$

$$A_1 = [1,1,1], \quad A_2 = [1,1,1]^T, \quad M = [W_1, W_2, W_3]$$

$$D_1 = D_2 = D_3 = \begin{vmatrix} g_1 & g_2 & g_3 & g_4 & T^{(b_i,h_j)} \\ g_5 & g_6 & g_7 & g_8 & T^{(b_1,h_j)} \\ g_9 & g_{10} & g_{11} & g_{12} & T_{local}^{(h_j)} \\ g_{13} & g_{14} & g_{15} & g_{16} & 1 \\ T^{(b_i,h_j)} & T^{(b_1,h_j)} & T_{local}^{(h_j)} & 1 & 0 \end{vmatrix}$$

$$W_1 = \begin{vmatrix} h_1 & g_2 & g_3 & g_4 & T^{(b_i,h_j)} \\ h_2 & g_6 & g_7 & g_8 & T^{(b_1,h_j)} \\ h_3 & g_{10} & g_{11} & g_{12} & T_{local}^{(h_j)} \\ h_4 & g_{14} & g_{15} & g_{16} & 1 \\ h_5 & T^{(b_1,h_j)} & T_{local}^{(h_j)} & 1 & 0 \end{vmatrix}$$

$$W_2 = \begin{vmatrix} g_1 & h_1 & g_3 & g_4 & T^{(b_i,h_j)} \\ g_5 & h_{26} & g_7 & g_8 & T^{(b_1,h_j)} \\ g_9 & h_3 & g_{11} & g_{12} & T_{local}^{(h_j)} \\ g_{13} & h_4 & g_{15} & g_{16} & 1 \\ T^{(b_i,h_j)} & h_5 & T_{local}^{(h_j)} & 1 & 0 \end{vmatrix}$$

$$W_3 = \begin{vmatrix} g_1 & g_2 & h_1 & g_4 & T^{(b_i,h_j)} \\ g_5 & g_6 & h_2 & g_8 & T^{(b_1,h_j)} \\ g_9 & g_{10} & h_3 & g_{12} & T_{local}^{(h_j)} \\ g_{13} & g_{14} & h_4 & g_{16} & 1 \\ T^{(b_i,h_j)} & T^{(b_i,h_j)} & h_5 & 1 & 0 \end{vmatrix}$$

**Theorem 5**: From **Theorem 4**, it has $Z^T H_1 Z X' = Z^T M$.

**Proof**: Let $Z$ be null-space matrix of the $A_1$, then we can have $A_1 Z = 0$, we assume the $[x_0, y_0, z_0]$ is feasible point that $\sum_{i=2}^{|BS|} x_{ij} + z_j + y_{1j} - 1 = 0$, so we conclude that $h_6 = 0$ and $A_1 U = 0$, and we can get that $U = ZX'$. Substituting this into the above equation we can get $Z^T H_1 Z X' = Z^T M$.

From **Theorem 5**, we can get the equation $Z^T H_1 Z X' = Z^T M$. Next, we use the conjugate gradient method and transform the equation into quadratic equation:

$$f(X) = \frac{1}{2} X^T Z^T H_1 Z X - Z^T M X \tag{33}$$

where $X^*$ can be solved as the solution points of AX=B.

Then we can get $X'$ by the conjugate gradient method, and denote $x = x + \rho \Delta x, y = y + \rho \Delta y$, and $z = z + \rho \Delta z$. We can use linear research technology to obtain the step $\rho$.

$$P8: \min \sum_{i=2}^{|BS|}\sum_{j=1}^{|H|} \alpha_{ij}(\hat{x}_{ij}^{k+1} - x_{ij}) + \sum_{i=2}^{|BS|}\sum_{j=1}^{|H|} \frac{\rho}{2}(\hat{x}_{ij}^{k+1} - x_{ij})^2 + \sum_{j=1}^{|H|} \beta_j \left(\hat{z}_j^{k+1} - z_j\right) + \sum_{j=1}^{|H|} \frac{\rho}{2}\left(\hat{z}_j^{k+1} - z_j\right)^2 + \sum_{j=1}^{|H|} \gamma_j \left(\hat{y}_{1j} - y_{1j}\right) + \sum_{j=1}^{|H|} \frac{\rho}{2}\left(\hat{y}_{1j} - y_{1j}\right)^2$$
$$-\omega \left(\sum_{i=1}^{|BS|}\sum_{j=1}^{|H|} \ln x_{ij} + \sum_{i=1}^{|BS|}\sum_{j=1}^{|H|} \ln(1-x_{ij}) + \sum_{j=1}^{|H|} \ln y_{1j} + \sum_{j=1}^{|H|} \ln(1-y_{1j}) + \sum_{j=1}^{|H|} \ln z_j + \sum_{j=1}^{|H|} \ln(1-z_j)\right) + \xi(\sum_{i=1}^{|BS|}\sum_{j=1}^{|H|} x_{ij}(1-x_{ij}) + \sum_{j=1}^{|H|} y_{1j}(1-y_{1j}) + \sum_{j=1}^{|H|} z_j(1-z_j)) \tag{23}$$

$$s.t \begin{cases} z_j T_{local}^{(s_j)} + \sum_{i=2}^{|BS|} x_{ij} T^{(b_i,h_j)} + y_{1j} T^{(b_1,h_j)} \leq t_j^{\max} \\ \sum_{i=2}^{|BS|} x_{ij} + z_j + y_{1j} = 1 \\ x_{ij} \in \{0,1\}, y_{1j} \in \{0,1\}, z_j \in \{0,1\} \\ i \in \{2,3...,|BS|\}, j \in \{1,2,3...,|H|\} \end{cases}$$

$$P9: \min G(x,y,z,m) = \sum_{i=2}^{|BS|}\sum_{j=1}^{|H|} \alpha_{ij}(\hat{x}_{ij}^{k+1} - x_{ij}) + \sum_{i=2}^{|BS|}\sum_{j=1}^{|H|} \frac{\rho}{2}(\hat{x}_{ij}^{k+1} - x_{ij})^2 + \sum_{j=1}^{|H|} \beta_j \left(\hat{z}_j^{k+1} - z_j\right) + \sum_{j=1}^{|H|} \frac{\rho}{2}\left(\hat{z}_j^{k+1} - z_j\right)^2 + \sum_{j=1}^{|H|} \gamma_j \left(\hat{y}_{1j} - y_{1j}\right) + \sum_{j=1}^{|H|} \frac{\rho}{2}\left(\hat{y}_{1j} - y_{1j}\right)^2 -$$
$$\omega \left(\sum_{i=1}^{|BS|}\sum_{j=1}^{|H|} \ln x_{ij} + \sum_{i=1}^{|BS|}\sum_{j=1}^{|H|} \ln(1-x_{ij}) + \sum_{j=1}^{|H|} \ln y_{1j} + \sum_{j=1}^{|H|} \ln(1-y_{1j}) + \sum_{j=1}^{|H|} \ln z_j + \sum_{j=1}^{|H|} \ln(1-z_j) + \sum_{j=1}^{|H|} \ln m_j\right) + \xi(\sum_{i=1}^{|BS|}\sum_{j=1}^{|H|} x_{ij}(1-x_{ij}) + \sum_{j=1}^{|H|} y_{1j}(1-y_{1j}) + \sum_{j=1}^{|H|} z_j(1-z_j)) \tag{26}$$

$$s.t \begin{cases} z_j T_{local}^{(s_j)} + \sum_{i=2}^{|BS|} x_{ij} T^{(b_i,h_j)} + y_{1j} T^{(b_1,h_j)} + m_j = t_j^{\max} \\ \sum_{i=2}^{|BS|} x_{ij} + z_j + y_{1j} = 1 \\ x_{ij} \in \{0,1\}, y_{1j} \in \{0,1\}, z_j \in \{0,1\} \\ i \in \{2,3...,|BS|\}, j \in \{1,2,3...,|H|\} \end{cases}$$

$$L(x,y,z,m,\upsilon,\varsigma) = \sum_{i=2}^{|BS|}\sum_{j=1}^{|H|} \alpha_{ij}(\hat{x}_{ij}^{k+1} - x_{ij}) + \sum_{i=2}^{|BS|}\sum_{j=1}^{|H|} \frac{\rho}{2}(\hat{x}_{ij}^{k+1} - x_{ij})^2 + \sum_{j=1}^{|H|} \beta_j \left(\hat{z}_j^{k+1} - z_j\right) + \sum_{j=1}^{|H|} \frac{\rho}{2}\left(\hat{z}_j^{k+1} - z_j\right)^2 + \sum_{j=1}^{|H|} \gamma_j \left(\hat{y}_{1j} - y_{1j}\right) + \sum_{j=1}^{|H|} \frac{\rho}{2}\left(\hat{y}_{1j} - y_{1j}\right)^2 -$$
$$\omega \left(\sum_{i=1}^{|BS|}\sum_{j=1}^{|H|} \ln x_{ij} + \sum_{i=1}^{|BS|}\sum_{j=1}^{|H|} \ln(1-x_{ij}) + \sum_{j=1}^{|H|} \ln y_{1j} + \sum_{j=1}^{|H|} \ln(1-y_{1j}) + \sum_{j=1}^{|H|} \ln z_j + \sum_{j=1}^{|H|} \ln(1-z_j) + \sum_{j=1}^{|H|} \ln m_j\right) + \xi(\sum_{i=1}^{|BS|}\sum_{j=1}^{|H|} x_{ij}(1-x_{ij}) + \sum_{j=1}^{|H|} y_{1j}(1-y_{1j}) + \sum_{j=1}^{|H|} z_j(1-z_j)) \tag{27}$$
$$+ \sum_{j=1}^{|H|} \upsilon_j (z_j T_{local}^{(s_j)} + \sum_{i=2}^{|BS|} x_{ij} T^{(b_i,h_j)} + y_{1j} T^{(b_1,h_j)} + m_j - t_j^{\max}) + \sum_{j=1}^{|H|} \varsigma_j \left(\sum_{i=2}^{|BS|} x_{ij} + z_j + y_{1j} - 1\right)$$

**Theorem 6**: Suppose that there exists $(x^*, y^*, z^*, m^*)$, then it has $\lim_{t \to \infty} Q(x(u_t), y(u_t), z(u_t), m(u_t)) = Q(x^*, y^*, z^*, m^*)$.

**Proof**: Please refer to the **Appendix D**.

V. PERFORMANCE EVALUATION

In this part, we evaluate the performance of the proposed task allocation mechanism. The parameters setting are shown in Table 1.

In Fig. 2, the number of tasks is 100, the number of BS is 6, the update step is 1.0, the computation capacity of SBS is 20GHz, and computation capacity of MBS is 100GHz. This figure clearly reveals the change of utility with the different parameters $\alpha$=0.2, $\alpha$=0.5, and $\alpha$=0.8 respectively. The utility experienced a significant decline, ending at stable value. From this figure, $\alpha$=0.8 ranked in the first, $\alpha$=0.5 was not far behind $\alpha$=0.8, while the figure for $\alpha$=0.2 was the smallest compared with other parameters. This is because that the influence of delay in utility overtakes that of energy consumption under this circumstance. So we can conclude that the utility will increase with increasing value of parameter $\alpha$.

Table 1. Simulation parameters setting.

| Parameter | Value |
|---|---|
| The transmit power of user | 0.1w |
| Input data size | 5 kbits-10kbits |
| Delay threshold | 15s-30s |
| Computational capability | 10-100GHz CPU cycles/s |
| The path loss exponent | 4 |
| The bandwidth | 20MHz |
| Computation workload | 18000 CPU cycles/bit |
| Computation efficiency coefficient | $10^{-26}$ |
| The power of noise | -172dBm/Hz |
| The transmit power of BS | 40 |
| BS coverage area | 200m*200m |
| $\alpha$ | 0.5 |
| $M^u_i$ | 100M |
| $\beta$ | 0.1 |
| $T$ | 1000 |

In Fig. 3, we verify the convergence of the improved ADMM algorithm in different step values. The number of tasks is 100, the number of BS is 6, the computational capacity of local terminal is 5GHz cycle/s, the computation capacity of SBS is 20GHz, the computation capacity of MBS is 100GHz, and the parameter $\alpha$ is 0.5. The rate of convergence in $\rho=1$ is slower than that in $\rho=1.2$. However, they are eventually converged in similar value. We can conclude that if the objective function is linear, we can get the optimal value.

Fig. 4 shows the changes in utility for different numbers of BSs among different numbers of tasks. The computational capacity is 5GHz cycle/s, and the step size $\rho$ is 1. The computational capacity of local terminal is 5GHz cycle/s, the computation capacity of SBS is 20GHz, and the computation capacity of MBS is 100GHz. The parameter $\alpha$ is 0.8. It is obvious that as the number of tasks increases, the utility will increase as well. With increasing the number of BSs (this means that the density of BS deployment increases), the distance between task and base station will decrease. From the above-mentioned analysis, we can know that the delay accounts for a large proportion of the total utility. Hence, if the distance decreases, it is obvious that utility will decrease as the density of base station increasing, as shown in Fig. 4.

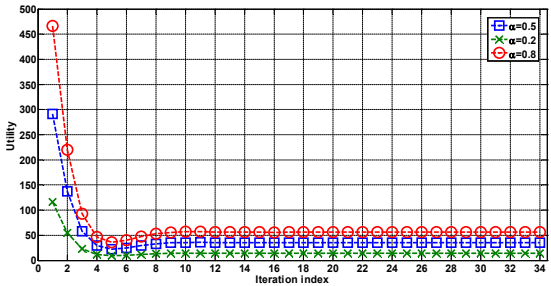

Fig. 2. Convergence progresses of the parallel multi-block ADMM-based algorithm (different weight factors $\alpha$ in **P1**).

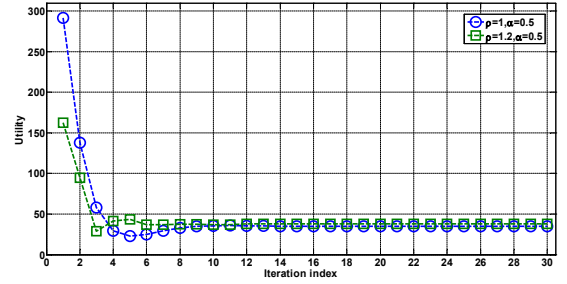

Fig.3. Convergence progresses of the parallel multi-block ADMM-based algorithm (different steps $\rho$).

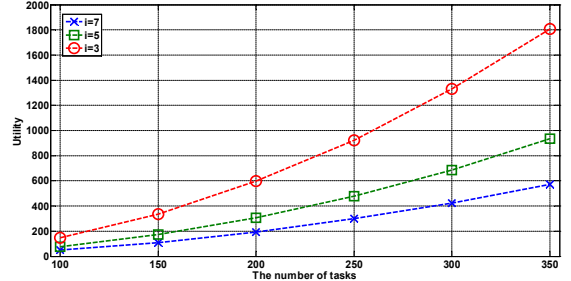

Fig. 4. The number of tasks vs. utility.

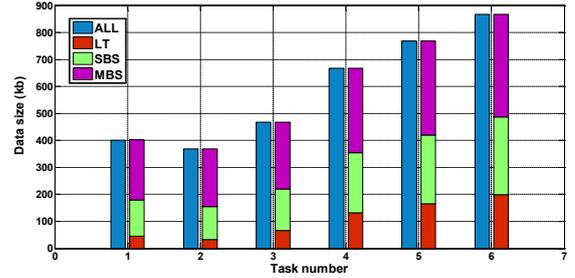

Fig. 5. The task number in three-tier framework vs. Data size.

In Fig. 5, we can draw several observations. The computational requirement of tasks is linearly related the size of tasks. When the size of task is smaller than 2.52kb, this task will be calculated in LT (local terminal). When the size of task is larger than 20Mb, our police will make task offload to MBS which can reduce energy consumption. When the size of tasks are 760.793kb, 867.93kb, 666.793kb, 467.93kb, 367.93kb, 400.3789 kb, we use three-tier model for task management, as shown in Fig. 5, which can make full use of resource among LT, SBS, and MBS, reducing delay and energy.

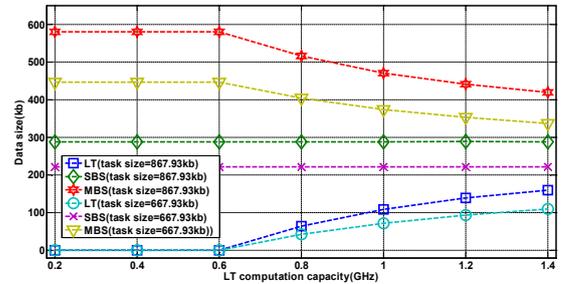

Fig. 6. LT computation capacity vs. Data size.

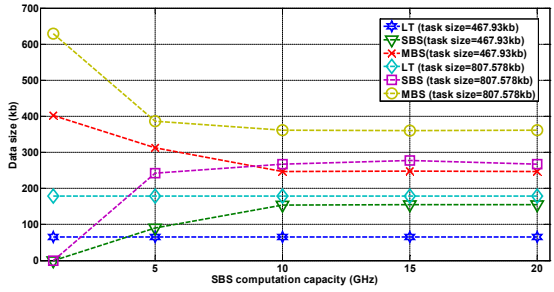

Fig. 7. SBS computation capacity vs. Data size.

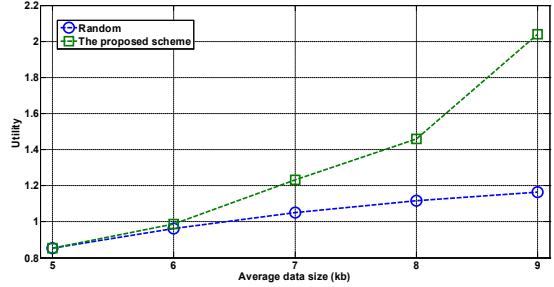

Fig. 8. Average data size vs. utility.

Fig. 6 shows the LT computation capacity for data size among LT, SBS and MBS. We can see that if LT computation capacity is less than 0.6 GHz, the tasks are not likely to process in LT; it will choose the there-tier model to process task. With the increase of LT computation capacity, we can see that the percentage of task which is processed in LT will increase, the reason is that the transition time is larger than the processing time, it is wildly know that the distance between the LT and MBS is larger than the distance between the LT and SBS, as the result the percentage of task which are processed in MBS will decrease.

Fig. 7 shows the SBS computation capacity for data size among LT, SBS and MBS. We can see that with the increase of SBS computation capacity, the percentage of task which is processed in SBS increase remarkably from 1GHz to 10GHz, and there was a slight increase from 10 GHz to 20GHz, the subtask which is processed in LT remained unchanged. The reason is that the processing time is larger than transition time for the larger data size. With the increase of SBS computation capacity, the subtask will choose to upload to SBS, when the size of the subtask enters into a certain value, it will choose to upload to MBS, which means that it will be useless to increase the SBS computation capacity. We can conclude that in our environment SBS computation capacity with 10GHz can meet our need.

Fig. 8 shows the average data size (1000 tasks) for utility between the random offloading scheme and our proposed scheme. The disparity between random offloading scheme and our proposed scheme will increase along with data size. In our proposed scheme, with the increase of data size, the task can be divided into several parts, and then it can make good use of the local, SBS and MBS computation capacity, and meanwhile can avoid congestion.

## VI. CONCLUSION

In this paper, we present a parallel optimal task allocation mechanism for a large-scale MEC system, which includes local computing model, three-tier model and MBS computation model. It can make full use of resource among MT, SBS, and MBS, featuring large-scale distributed optimization and efficient task allocation mechanism, to reach intelligent and efficient resource management in MEC. In the proposed mechanism, we formulate the task allocation as nonlinear 0-1 integer programming problem, which can use Logarithmic Smoothing by modifying the inequality constrain as equality constrain in decision variables and smoothing function. In the meantime, we exploit regulation linear method to make the optimal function to be linear, guaranteeing its convergence and it can run in a parallel manner. For the solutions, we exploit the conjugate gradient and Newton methods, and CBGP method to solve the optimal problem and sub-optimal problem, respectively, in which the global and local variables can update simultaneously. The simulation results show that the proposed mechanism can guarantee the convergence and effectively reduce delay and energy consumption in a large-scale MEC system.

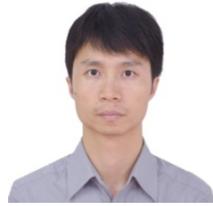
**Xiaoxiong Zhong** (S'12-M'16) received his Ph.D degree in Computer Science and Technology from Harbin Institute of Technology, China, in 2015. He was a Postdoctoral Research Fellow with Tsinghua University, from 2016 to 2018. He is currently an assistant professor with the Cyberspace Security Research Center, Peng Cheng Laboratory, Shenzhen, China. His general research interests include network protocol design and analysis, data transmission and data analysis in internet of things and edge computing.

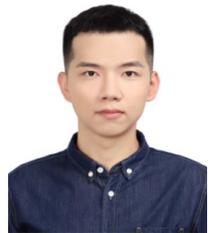
**Xinghan Wang** received his B.S. degree in Computer Science and Technology from Taiyuan University of Technology, China, in 2016. He is currently pursuing the M.S. degree in the School of Computer Science and Information Security at Guilin University of Electronic Technology, China. His research interest includes network protocol design and analysis, reinforcement learning and edge computing.

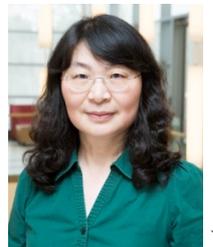
**Yuanyuan Yang** received the BEng and MS degrees in computer science and engineering from Tsinghua University and the MSE and PhD degrees in computer science from Johns Hopkins University. She is a SUNY Distinguished Professor of computer engineering and computer science and the Associate Dean for Academic Affairs in the College of Engineering and Applied Sciences at Stony Brook University, New York. Her research interests include wireless networks, data center networks and cloud computing. She has published over 380 papers in major journals and refereed conference proceedings and holds seven US patents in these areas. She is currently the Associate Editor-in-Chief for IEEE Transactions on Cloud Computing and an Associate Editor for ACM Computing Surveys. She has served as an Associate Editor-in-Chief and Associated Editor for IEEE Transactions on Computers and Associate Editor for IEEE Transactions on Parallel and Distributed Systems. She has also served as a general chair, program chair, or vice chair for several major conferences and a program committee member for numerous conferences. She is an IEEE Fellow.

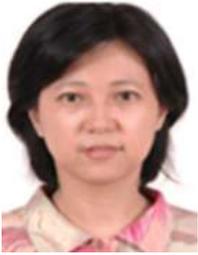 **Yang Qin** (S'98-M'01-SM'06) received her B.S (with first class honors) in Computer Science at Southwest Jiaotong University (China), in 1989, M.S in Computer Science at Huazhong University of Science & Technology, Wuhan, Hubei, in 1992, and Ph.D in Computer Science, Hong Kong University of Science & Technology, Kowloon, Hong Kong at November of 1999. From 1999 to 2000, she has visited the Washington State University as a Postdoc, USA. From 2000 to 2008, she is an assistant professor Nanyang Technological University, Singapore. Currently, she is an associate professor in the Department of Computer Science and Technology, Harbin Institute of Technology (Shenzhen), China. Her research interests are in the areas of wireless networks, mobile computing, cross-layer design, QoS of routing and scheduling, high speed optical networks and so forth. She is a senior member of IEEE.

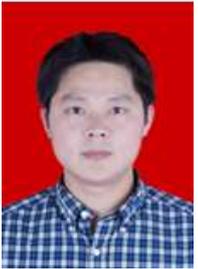 **Xiaoke Ma** received the PhD degree in computer science from Xidian University, in 2012. He is a professor with the School of Computer Science and Technology, Xidian University, China. His research interests include machine learning, optimization, data mining, and bioinformatics. He publishes more than 40 papers in the peer-reviewed international journals, such as the IEEE Transactions on Knowledge and Data Engineering, the IEEE Transactions on Cybnetics, Pattern Recognition, Bioinformatics, PLoS Computational Biology, Cell Stem Cell, Nuclear Acids Research, and the IEEE/ACM Transactions on Computational Biology and Bioinformatics.

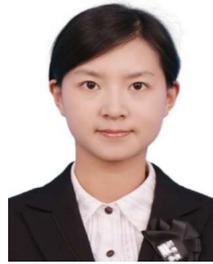 **Tingting Yang** (M'13) received the Ph.D. degrees from Dalian Maritime University, China, in 2010. She is currently a professor in the School of Electrical Engineering and Intelligentization, Dongguan University of Technology, China. Since September 2012, she has been a visiting scholar at the Broadband Communications Research (BBCR) Lab at the Department of Electrical and Computer Engineering, University of Waterloo, Canada. Her research interests are in the areas of maritime wideband communication networks, DTN networks, and green wireless communication. She serves as the associate Editor-in-Chief of the IET Communications, as well as the advisory editor for SpringerPlus. She also serves as a TPC Member for IEEE ICC'14, ICC'15.

# Appendices for the paper
# "A Parallel Optimal Task Allocation Mechanism for Large-Scale Mobile Edge Computing"

## APPENDIX A
PROOF OF THEOREM 1

**Proof**: When $x_{ij}=1$, from the first constraint we can get $R_{ij} \geq 1$, and from the fourth constraint we can get $R_{ij} \geq r^{(b_i,h_j)}$, since the variable $h^{(b_i,h_j)}$ meets that $0 \leq h^{(b_i,h_j)} \leq 1$, so we can get $r^{(b_i,h_j)} \geq 1$. The first and fourth constraints we can get $R_{ij} \geq r^{(b_i,h_j)}$. The second constraints we can get $R_{ij} \leq \frac{1}{h_{min}^{(b_i,h_j)}}$, the third constraints we can get $R_{ij} \leq r^{(b_i,h_j)}$. Since the variable meets that $r^{(b_i,h_j)} \leq \frac{1}{h_{min}^{(b_i,h_j)}}$, the second and third constraints we can get $R_{ij} \leq r^{(b_i,h_j)}$. So when $x_{ij}=1$, $r^{(b_i,h_j)} \leq R_{ij} \leq r^{(b_i,h_j)}$, as a result, we can get $R_{ij}=r^{(b_i,h_j)}$. When $x_{ij}=0$, from the first constraint we can get $R_{ij} \geq 0$, from the fourth constraint we can get $R_{ij} \geq r^{(b_i,h_j)} - \frac{1}{h_{min}^{(b_i,h_j)}}$. Since $r^{(b_i,h_j)} \leq \frac{1}{h_{min}^{(b_i,h_j)}}$, from the first and fourth constraints we can get $R_{ij} \geq 0$. From the second constraint we can get $R_{ij} \leq 0$, and from the third constraint we can get $R_{ij} \leq r^{(b_i,h_j)} - 1$. Since $r^{(b_i,h_j)} \geq 1$, from the second and third constraints we can get $R_{ij} \leq 0$, So when $x_{ij}=0$, $0 \leq R_{ij} \leq 0$. As a result, we can get $R_{ij}=0$.

Hence, the Equitation (34) can be replaced by Equitation (35). And then the objective function in **P2** can be changed as a linear function.

$$\xi_i = \left\{ \begin{array}{l} \hat{x}_{ij} \\ h^{(b_i,h_j)} \\ \hat{y}_{1j} \\ \hat{z}_j \end{array} \middle| \begin{array}{l} \hat{y}_{1j} \in \{0,1\} \\ \hat{x}_{ij} \in \{0,1\} \\ 0 < h^{(b_i,h_j)} \leq 1 \\ \sum_{j=1}^{|H|} h^{(b_i,h_j)} \hat{x}_{ij} \leq 1 \\ \hat{z}_j \in \{0,1\} \end{array} \right\} \quad (34)$$

$$\xi_i = \left\{ \begin{array}{l} \hat{x}_{ij} \\ \frac{1}{r^{(b_i,h_j)}} \\ \hat{y}_{1j} \\ \hat{z}_j \\ \hat{R}_{ij} \end{array} \middle| \begin{array}{l} \hat{x}_{ij} \in \{0,1\} \\ \sum_{j=1}^{|H|} \frac{1}{r^{(b_i,h_j)}} \leq 1 \\ \hat{z}_j \in \{0,1\} \\ \hat{x}_{ij} \leq R_{ij} \leq \hat{x}_{ij} \frac{1}{h_{min}^{(b_i,h_j)}} \\ R_{ij} \leq r^{(b_i,h_j)} + \hat{x}_{ij} - 1 \\ R_{ij} \geq r^{(b_i,h_j)} + \hat{x}_{ij} \frac{1}{h_{min}^{(b_i,h_j)}} - \frac{1}{h_{min}^{(b_i,h_j)}} \\ \hat{y} \in \{0,1\} \\ i \in \{2,3...,|BS|\}, j \in \{1,2,3...,|H|\} \end{array} \right\} \quad (35)$$

## APPENDIX B
PROOF OF THEOREM 2

**Proof**: $x_{ij}^{k+1}$ is better than $x_{ij}^k$, so we can conclude that $x_{ij}^{k+1} - (x_{ij}^{k+1})^2 < x_{ij}^k - (x_{ij}^k)^2$.

$$\begin{aligned} & x_{ij}^{k+1} - (x_{ij}^{k+1})^2 \\ & \leq x_{ij}^{k+1} - (x_{ij}^k)^2 - \langle \nabla(x_{ij}^k)^2, x_{ij}^{k+1} - x_{ij}^k \rangle \\ & \leq x_{ij}^k - (x_{ij}^k)^2 - \langle \nabla(x_{ij}^k)^2, x_{ij}^k - x_{ij}^k \rangle \\ & = x_{ij}^k - (x_{ij}^k)^2 \end{aligned}$$

which means that
$$x_{ij}^{k+1} - (x_{ij}^k)^2 - \langle \nabla(x_{ij}^k)^2, x_{ij}^{k+1} - x_{ij}^k \rangle \leq x_{ij}^k - (x_{ij}^k)^2 - \langle \nabla(x_{ij}^k)^2, x_{ij}^k - x_{ij}^k \rangle.$$

Hence, $\sum_{i=1}^{|BS|} \sum_{j=1}^{|H|} x_{ij} - x_{ij}^2$ can be replaced by $x_{ij}^{k+1} - (x_{ij}^k)^2 - \langle \nabla(x_{ij}^k)^2, x_{ij}^{k+1} - x_{ij}^k \rangle$. And then if we add the equation into **P3**, then we can obtain **P7**.

## APPENDIX C
PROOF OF THEOREM 3

**Proof**: Denote $\nabla y(x)$ as the partial derivative of $y$ corresponding to the variable $x$, define $P_\Omega(x) = \arg\min_{y \in \Omega} \|x - y\|^2$. So we can conclude that:

Given $\hat{x}_{ij}$, $c_{0j}$, $c_{ij}$, $c_{1j}$ we update $R_{ij}$

$$R_{ij}^{(k+1)} = P_{\Omega_{R_{ij}}}(R_{ij}^{(k)} - \varepsilon_{ij}\nabla Q(R_{ij}^{(k)})) \quad (36)$$

where $\Omega_{R_{ij}}$ is the bounded domain constrained of variables.

We update the variables $c_{0j}$, $c_{1j}$ and $c_{ij}$ sequentially according to CBGP method. In updating, the objective function is convex and the constraints are linear, so we use subgradient to update the variables.

$$L(\hat{x}_{ij}, \varepsilon_{ij}, \phi_{ij}, \varpi_{ij}, \theta_{ij}, \vartheta_{ij}, \sigma_{ij}) = Q(\hat{x}_{ij}, R_{ij}, c_{1j}, c_{ij}, c_{0j}) - \varepsilon_{ij}x_{ij} + \phi_{ij}(x_{ij} - 1) + \varpi_{ij}(\hat{x}_{ij} - R_{ij}) + \theta_{ij}(R_{ij} - \hat{x}_{ij}\frac{1}{h_{\min}^{(b_i, h_j)}})$$
$$+ \vartheta_{ij}(R_{ij} - (r^{(b_i, h_j)} + \hat{x}_{ij} - 1)) + \sigma_{ij}(r^{(b_i, h_j)} + \hat{x}_{ij}\frac{1}{h_{\min}^{(b_i, h_j)}} - \frac{1}{h_{\min}^{(b_i, h_j)}} - R_{ij}) \quad (37)$$

$$x_{ij}^{k+1*} = -\frac{\alpha T^{(h_i, c_{ij})} + \varpi_{ij} + \theta_{ij}(-\frac{1}{h_{\min}^{(b_i, h_j)}}) - \vartheta_{ij} + \sigma_{ij}(\frac{1}{h_{\min}^{(b_i, h_j)}})}{\rho} - \frac{(1-\alpha)E^{(h_i, h_j)} - \rho x_{ij}^k + \delta_{ij}(1 - 2x_{ij}^k) - \varepsilon_{ij} + \phi_{ij}}{\rho} \quad (39)$$

We utilize the Lagrangian dual decomposition method to solve the above problem. The Lagrangian function can be expressed by (37).

Then Lagrangian dual problem is expressed as:

$$\min L(\hat{x}_{ij}, \varepsilon_{ij}, \phi_{ij}, \varpi_{ij}, \theta_{ij}, \vartheta_{ij}, \sigma_{ij}) = \max g(\varepsilon_{ij}, \phi_{ij}, \varpi_{ij}, \theta_{ij}, \vartheta_{ij}, \sigma_{ij})$$
$$s.t. \quad \varepsilon_{ij} \geq 0, \phi_{ij} \geq 0, \varpi_{ij} \geq 0, \theta_{ij} \geq 0, \vartheta_{ij} \geq 0, \sigma_{ij} \geq 0 \quad (38)$$

Base on KKT conditions, the optimal can be expressed as (39).

Next, we use the sub-gradient method to solve the dual problem. The dual variables are update as (43):

$$\begin{cases} \varepsilon_{ij}^{k+1} = \left[\varepsilon_{ij}^k + \zeta_{ij} \times \nabla g(\varepsilon_{ij}^k)\right]^+ \\ \phi_{ij}^{k+1} = \left[\phi_{ij} + \psi_{ij} \times \nabla g(\phi_{ij}^k)\right]^+ \\ \varpi_{ij}^{k+1} = \left[\varpi_{ij}^k + \xi_{ij} \times \nabla g(\varpi_{ij}^k)\right]^+ \\ \theta_{ij}^{k+1} = \left[\theta_{ij}^k + \omega_{ij} \times \nabla g(\theta_{ij}^k)\right]^+ \\ \vartheta_{ij}^{k+1} = \left[\vartheta_{ij}^k + o_{ij} \times \nabla g(\vartheta_{ij}^k)\right]^+ \\ \sigma_{ij}^{k+1} = \left[\sigma_{ij}^k + v_{ij} \times \nabla g(\sigma_{ij}^k)\right]^+ \end{cases} \quad (40)$$

where $\zeta_{ij} > 0, \psi_{ij} > 0, \xi_{ij} > 0, \omega_{ij} > 0, o_{ij} > 0, v_{ij} > 0$, and $|x|^+$ means $x \in \max\{0, x\}$.

## APPENDIX D
PROOF OF THEOREM 6

**Proof**: We can simply get following equation:

$$\begin{cases} \nabla_x G(x^*, y^*, z^*, m^*) + T^{(b_i, h_j)}v_j^* + \varsigma_j^* = 0 \\ \nabla_y G(x^*, y^*, z^*, m^*) + T^{(b_1, h_j)}v_j^* + \varsigma_j^* = 0 \\ \nabla_z G(x^*, y^*, z^*, m^*) + T_{local}^{(h_j)}v_j^* + \varsigma_j^* = 0 \\ \nabla_m G(x^*, y^*, z^*, m^*) + v_j^* = 0 \\ z_j^* T_{local}^{(h_j)} + \sum_{i=2}^{|BS|} x_{ij}^* T^{(b_i, h_j)} + y_{1j}^* T^{(b_1, h_j)} + m_j^* = t_j^{\max} \\ \sum_{i=2}^{|BS|} x_{ij}^* + z_j^* + y_{1j}^* = 1 \end{cases} \quad (41)$$

and for each $t$ we have

$$\begin{cases} \nabla_x G(x(u_t), y(u_t), z(u_t), m(u_t)) + T^{(b_i, h_j)}v_j(u_t) + \varsigma_j(u_t) = 0 \\ \nabla_y G(x(u_t), y(u_t), z(u_t), m(u_t)) + T^{(b_1, h_j)}v_j(u_t) + \varsigma_j(u_t) = 0 \\ \nabla_z G(x(u_t), y(u_t), z(u_t), m(u_t)) + T_{local}^{(h_j)}v_j(u_t) + \varsigma_j(u_t) = 0 \\ \nabla_m G(x(u_t), y(u_t), z(u_t), m(u_t)) + v_j(u_t) = 0 \\ z_j(u_t)T_{local}^{(h_j)} + \sum_{i=2}^{|BS|} x_{ij}(u_t)T^{(b_i, h_j)} + y_{1j}(u_t)T^{(b_1, h_j)} + m_j(u_t) = t_j^{\max} \\ \sum_{i=2}^{|BS|} x_{ij}(u_t) + z_j(u_t) + y_{1j}(u_t) = 1 \end{cases} \quad (42)$$

From above equation we can conclude that:

$$\begin{bmatrix} T_{local}^{(h_j)} & T^{(b_i, h_j)} & T^{(b_1, h_j)} & 1 \\ & & & \\ & & & \\ 1 & 1 & 1 & 0 \end{bmatrix} \begin{bmatrix} x(u_t) - x^* \\ y(u_t) - y^* \\ z(u_t) - z^* \\ m(u_t) - m^* \end{bmatrix} = \begin{bmatrix} 0 \\ \\ \\ 0 \end{bmatrix}$$
$$\quad (43)$$

$$\begin{bmatrix} H_1 \\ H_2 \\ H_3 \\ H_4 \end{bmatrix} + \begin{bmatrix} T^{(b_i,h_j)} & 1 \\ T^{(b_1,h_j)} & 1 \\ T_{local}^{(h_j)} & 1 \\ 1 & 0 \end{bmatrix} \begin{bmatrix} \upsilon_j(u_t)-\upsilon_j^* \\ \\ \varsigma_j(u_t)-\varsigma_j^* \end{bmatrix} = \begin{bmatrix} 0 \\ 0 \\ 0 \\ 0 \end{bmatrix} \quad (44)$$

where

$$H_1 = \nabla_x G(x(u_t), y(u_t), z(u_t), m(u_t)) - \nabla_x G(x^*, y^*, z^*, m^*)$$
$$H_2 = \nabla_y G(x(u_t), y(u_t), z(u_t), m(u_t)) - \nabla_y G(x^*, y^*, z^*, m^*)$$
$$H_3 = \nabla_z G(x(u_t), y(u_t), z(u_t), m(u_t)) - \nabla_z G(x^*, y^*, z^*, m^*)$$
$$H_4 = \nabla_m G(x(u_t), y(u_t), z(u_t), m(u_t)) - \nabla_m G(x^*, y^*, z^*, m^*)$$
$$(45)$$

Multiplying the both sides $\begin{bmatrix} x(u_t)-x^* \\ y(u_t)-y^* \\ z(u_t)-z^* \\ m(u_t)-m^* \end{bmatrix}^T$, we can conclude that:

$$\begin{bmatrix} x(u_t)-x^* \\ y(u_t)-y^* \\ z(u_t)-z^* \\ m(u_t)-m^* \end{bmatrix}^T \begin{bmatrix} H_1 \\ H_2 \\ H_3 \\ H_4 \end{bmatrix} = \mathbf{0} \quad (46)$$

Since $H_1, H_2, H_3, H_4$ are linearly independent. So we can conclude that

$$\begin{cases} x(u_t)-x^* = 0 \\ y(u_t)-y^* = 0 \\ z(u_t)-z^* = 0 \\ m(u_t)-m^* = 0 \end{cases} \quad (47)$$